\documentstyle[aps,preprint,prb,graphicx,tighten]{revtex}
\begin{document}
\title{Flexibility of $\beta$-sheets: Principal-component
analysis of database protein structures}
\author{Eldon G. Emberly$^1$, Ranjan Mukhopadhyay$^2$, Chao Tang$^2$,
and Ned S. Wingreen$^{2,*}$ }
\address{
$^1$Center for Studies in Physics and Biology, The
Rockefeller University, \\1280 York Ave., New York, NY 10021\\
$^2$NEC Laboratories America, Inc.,\\ 4 Independence Way,\\ Princeton,
NJ 08540\\ $^*$ corresponding author: email: wingreen@nec-labs.com
\\ P:~(609)~951-2654 F:~(609)~951-2496\\
}
\date{\today}
\maketitle

keywords: beta-sheet, flexibility, principal-component analysis,
secondary structure

\renewcommand{\thefootnote}{\fnsymbol{footnote}}
\begin{abstract}
Protein folds are built primarily from the packing together of two
types of structures: $\alpha$-helices and $\beta$-sheets. Neither
structure is rigid, and the flexibility of helices and sheets is often
important in determining the final fold ({\it e.g.}, coiled coils and
$\beta$-barrels). Recent work has quantified the flexibility of
$\alpha$-helices using a principal-component analysis (PCA) of
database helical structures (Emberly, 2003). Here, we extend the
analysis to $\beta$-sheet flexibility using PCA on a database of
$\beta$-sheet structures. For sheets of varying dimension and
geometry, we find two dominant modes of flexibility: twist and
bend. The distributions of amplitudes for these modes are found to be
Gaussian and independent, suggesting that the PCA twist and bend modes
can be identified as the soft elastic normal modes of sheets. We
consider the scaling of mode eigenvalues with sheet size and find that
parallel $\beta$-sheets are more rigid than anti-parallel sheets over
the entire range studied. Lastly, we discuss the application of our
PCA results to modeling and design of $\beta$-sheet proteins.

\end{abstract}
\section{Introduction}
Most protein folds can be viewed as compact packings of a fixed set of
secondary-structural elements\cite{Richardson81,Chothia77}:
$\alpha$-helices and $\beta$-sheets.  It can be reasoned that the
formation of these elements greatly simplifies the folding free-energy
landscape by reducing the number of degrees of freedom. As a first
approximation, helices and sheets can be considered as rigid objects,
possessing only six degrees of freedom each (three translations and
three rotations).  However, most helices and sheets display some
amount of bending in a protein's final fold. Understanding to what
extent these elements are flexible, and which are their dominant
degrees of freedom, will help to further our understanding of how
proteins fold, and even how they function.

A number of studies have extracted the flexible motions
of biological molecules using normal-mode and principal-component
analysis (PCA).\cite{Olson,Kidera,Diamond,Faure,Tirion,Haliloulu,Krebs,Travers,Emberly}
A recent PCA analysis of $\alpha$-helices from a structural database
revealed three ``soft'' modes: two degenerate bend modes and a twist 
mode.\cite{Emberly} For all but the longest helices, these three
modes were sufficient to describe the deformations observed in
real structures.

Arguably, a quantitative understanding of flexibility is 
more important for $\beta$-sheets than for $\alpha$-helices.
In natural structures, sheets display a variety of highly distorted and 
bent shapes, {\it e.g.} $\beta$-barrels and twisted sheets, while
helices are generally much less distorted. What are the dominant 
collective motions of $\beta$-sheets and how do they depend on
the sheet's size? Also how does sheet geometry affect the flexibility
of a sheet? The ``geometry'' of a $\beta$-sheet is the amino-terminal
to carboxyl-terminal orientation of the various strands making up the sheet.
Most sheets fall into one of two geometries - parallel, where all the
strands are oriented in the same direction, or anti-parallel, where
strands alternate direction. The geometry dictates the hydrogen-bonding 
pattern within the sheet and hence plays a role in determining the 
sheet's flexibility.

Here, we report a principal-component analysis of the flexibility of
parallel and anti-parallel $\beta$-sheets from the Protein Data Bank. The 
sheets considered range in size from 3 to 6 strands, with 
3 to 6 residues per strand. For both parallel and anti-parallel
sheets, we find two dominant modes of flexibility: twisting about an
in-plane axis that is perpendicular to the strand orientation, and
bending of this same axis. The distributions of amplitudes
for these two modes are independent Gaussians. Thus, the principal-component 
modes can be interpreted as dynamical normal modes of an elastic object. 
Motivated by this interpretation, we consider the scaling of mode
eigenvalues (variances of amplitudes) with sheet size, and compare
to predictions of a simple elastic model. For all sizes considered, 
parallel sheets are more rigid than anti-parallel sheets.

Recently, $\beta$-sheet structures have been characterized in detail
by Ho and Curmi.\cite{Ho02}. This database study focused on average
properties of sheets, including twist, shear, and hydrogen bonding.
In contrast, our PCA analysis of $\beta$-sheets provides a
characterization of the {\it flexibility} of sheets about their
average structures.  Possible applications of the results reported
here on sheet flexibility include improved parameterization of
force-field models and inclusion of sheet elastic energies in
$\beta$-protein design.

\section{Results}
\subsection{Principal-component analysis of database $\beta$-sheets}

The first step in analyzing the flexibility of $\beta$-sheets was to
obtain a representative set of structures. Following the procedure
described in Methods, we were able to create a database of $3516$
representative $\beta$-sheets of different geometries.  In
Fig.~\ref{fig1} we show the distribution of geometries (parallel,
anti-parallel, etc.)  for sheets consisting of at least 6 strands. For
sheets that have more than 6 strands, we consider all the 6 stranded
sub-sheets (e.g. an 8 stranded sheet would contain three 6 strand
sheets). The strand directions are labeled `u' for up and `d' for
down, as if the sheet were lying flat on the page.  By convention, the
first strand is always oriented in the up direction, {\it i.e.} with
the amino-terminal at the bottom of the page and the carboxyl-terminal
at the top of the page. Since each of the two outermost strands of a
sheet can be equally well regarded as the first strand, the
distribution of geometries in Fig.~\ref{fig1} includes every sheet
twice.  We find that the distribution of geometries is highly
non-uniform, with the most frequent types being parallel (uuuuuu) and
anti-parallel (ududud). Because other geometries occur so
infrequently, we focus the rest of the analysis on parallel and
anti-parallel sheets.

Sets of defect-free $\beta$-sheets of a given class were extracted
from the 3516 representative $\beta$-sheets. Sheets are of the same
class if they have the same size, $S$ strands each of length $L$,
geometry, and pleatedness or corrugation (see Methods). To quantify
flexibility, we performed a structural principal-component analysis
(PCA) for each class of sheets, to identify the dominant collective
fluctuations around the mean structure.  To implement the PCA, we
first computed the mean structure for each class of sheets via an
iterative procedure.  Starting with a randomly chosen sheet of the
desired class, we aligned to it all other sheets of the same class. To
align two sheets, we minimized the coordinate root mean square (crms)
distance between their corresponding C$_\alpha$ atoms. A mean
structure was then obtained by averaging the position of each
C$_\alpha$ atom over all the aligned structures. This procedure was
iterated, each time using the new mean structure as the basis for
alignment, until the mean structure converged to within $10^{-4}$
\AA/residue. An example of a subset of structures aligned to the
converged mean structure is shown in Fig.~\ref{fig2}. For
anti-parallel sheets we find that the average structure conforms to
the sheared structure discussed in Ho and Curmi\cite{Ho02}. They also
showed that parallel sheets are less sheared and we find this to be
the case for our average parallel sheets.

The second step in the principal-component analysis was to compute the
structural covariance matrix for each class of sheet. A covariance
matrix measures the correlation of the variation from the mean
for each pair of coordinates. In our case, there were $3SL$ coordinates --
3 spatial directions for each of $S \times L$ C$_\alpha$ atoms.
Consequently, the covariance matrix was a $3SL \times 3SL$ matrix,
with elements $i,j$ defined as
\begin{equation}
C_{i,j} = \frac{1}{N-1} \sum_{m=1}^{N} (x_{mi} - \langle x_i\rangle)
(x_{mj} - \langle x_j\rangle),
\end{equation}
where $N$ is the number of sheets in the given class, 
$x_{mi}$ is the $i^{th}$ coordinate of the $m^{th}$ structure, and
$\langle x_i\rangle$ is the $i^{th}$ coordinate of the mean structure.

To complete the principal-component analysis, we computed the
eigenvalues $\{\lambda_q\}$ and eigenvectors $\{\vec{v}_q\}$ of the
covariance matrix for each class of sheets (available as Supplementary
Material).  The largest eigenvalues and corresponding eigenvectors
represent the directions for which the data has the largest
variance. These directions are the ``soft'' modes of the sheets, {\it
i.e.} those collective deformations that appear with largest amplitude
in the data set.  Figure~\ref{fig3} shows the top 10 eigenvalues for
anti-parallel sheets of size $S=4$ and $L=5$, as shown in
Fig.~\ref{fig2}. Each eigenvalue is given in units of \AA$^2$ and
measures the variance of the distribution for a particular mode.  Two
dominant eigenvalues are evident in Fig.~\ref{fig3}.  The first mode
is primarily a twist of the sheet about the in-plane axis
perpendicular to the strand orientation (Fig.~\ref{fig4}(a)). The
second mode is primarily a bend of the sheet along the same axis
(Fig.~\ref{fig4}(b)).

For all the classes of sheets considered, we found two dominant soft
modes. The eigenvalues ({\it i.e.} variances) of these modes are shown
for different sheet sizes and geometries in
Figs.~\ref{fig5}~and~\ref{fig6}. Figure~\ref{fig5} shows the scaling
of the eigenvalues with the number of strands $S$ for sheets of fixed
strand length. The bend-mode eigenvalues increase approximately as
$S^4$, while the twist-mode eigenvalues increase more slowly with $S$
(fits to $S^4$ for the anti-parallel bend modes are shown).  As a
result of this difference in scaling, the eigenvalues for the twist
and bend modes cross with increasing number of strands. For five or
more strands, the eigenvalue for the bend mode becomes the larger,
implying greater deformations of the sheet by bending compared to
twisting.  Figure~\ref{fig6} shows the scaling of the eigenvalues with
the strand length $L$ for sheets with a fixed number of strands.  The
eigenvalues generally increase with strand length, scaling roughly as
$L$ or $L^2$. The scaling behavior expected for pure bend and twist
modes is discussed in the next section.

A feature that emerges from the scaling graphs is that the twist-mode
and bend-mode eigenvalues are almost always larger for anti-parallel 
sheets than for parallel sheets. Since these two modes dominate
deviations from the mean structure, this implies that total deformations 
are typically larger for anti-parallel sheets than for parallel sheets.

Next, we consider the actual distributions of amplitudes for the
dominant twist and bend modes. The displacement of a given sheet from
the mean structure, $\delta\vec{x} = \vec{x} - \langle\vec{x}\rangle$,
can be expanded in terms of the PCA eigenvectors as $\delta\vec{x} =
\sum_q a_q \vec{v}_q$.  The amplitude $a_q$ is given by the projection
of the displacement vector $\delta\vec{x}$ onto mode $q$.
Figures~\ref{fig7}(a and b) show the distributions of projections onto
twist and bend for the $1454$ anti-parallel sheets with $S=4$ and
$L=5$ (cf. Fig.~\ref{fig2}).  The two distributions can be fit well by
Gaussians, with the variances of the Gaussians, $9.6985$ \AA$^2$ for
twist and $4.3655$ \AA$^2$ for bend, close to the exact variances
given by the mode eigenvalues, $7.7876$ \AA$^2$ for bend and $4.0044$
\AA$^2$ for twist.  By construction, all PCA modes are uncorrelated to
lowest order, {\it i.e.} $\langle a_q a_{q'} \rangle = 0$, for $q \neq
q'$.  To look for possible higher-order correlations, we made a
scatter plot of the amplitudes for the two dominant modes, as shown in
Fig.~\ref{fig7}(c). The distributions of points is roughly
ellipsoidal, indicating that there are no strong higher-order
correlations between modes. Similar independent Gaussian distributions
were obtained for all classes of sheets for which PCA analysis was
performed.

The Gaussian distributions of mode amplitudes and the lack of
higher-order correlations between modes suggest that the PCA modes can
be interpreted as the {\it dynamical} modes of a $\beta$-sheet. This
is consistent with previous results for $\alpha$-helices showing the
near identity between PCA modes obtained from static structures and
the elastic normal modes of a model helix.\cite{Emberly} For small
amplitude motions, elastic normal modes at equilibrium have
independent Gaussian distributions $P(a_q)$ determined by Boltzmann
weights
\begin{equation}
P(a_q) \sim e^{-E(a_q)/k_BT} \sim e^{-c_q a_q^2/2k_BT},
\label{boltzmann}
\end{equation}
where $E(a_q) = c_q a_q^2 / 2$ is the deformation energy
as a function of the mode amplitude $a_q$, and $c_q$ is 
the spring constant for the mode. The ``soft'' modes have 
the smallest spring constants, and therefore the broadest 
spread of amplitudes.

\subsection{Scaling of the PCA modes}

Guided by the interpretation of the PCA modes as the elastic normal
modes of a sheet, we consider the scaling with sheet size of the mode
eigenvalues.  Let us first consider the dominant bend mode, as shown
in Fig.~\ref{fig4}(b).  For a uniform bend of the in-plane axis
perpendicular to the strand orientation, the displacement of the
strand at position $x$ along this axis goes as $\delta z \simeq
x^{2}/R$, where $R$ is the radius of curvature.  The bending
eigenvalue is given by $\lambda_{\rm bend} = \langle|\delta {\vec z}
\cdot \vec{v}|^{2} \rangle$ where $\vec{v}$ is the normalized
eigenvector. It follows that
\begin{equation}
\lambda_{\rm bend} \sim L S^{5} \langle
\frac{1}{R^2} \rangle,
\end{equation}
where $L$ is the strand length and $S$ is the number of strands.
At thermal equilibrium, each normal mode has $k_{B}T/2$ of potential
energy. For the bend mode, this energy would be put into curvature
of the axis:
\begin{equation}
\frac{1}{2} k_{B} T = \frac{1}{2} \kappa L S  \langle
\frac{1}{R^2} \rangle,
\end{equation}
where $\kappa$ is the bend stiffness per unit length, indicating
\begin{equation}
\lambda_{\rm bend} \sim \frac{k_{b}T}{\kappa} S^{4}.
\end{equation}
Thus the eigenvalue for a pure bend mode would scale as $S^{4}$ and 
would be independent of $L$. In Fig.~\ref{fig5}, the predicted $S^4$ 
scaling of the bend-mode eigenvalue is seen for both parallel and 
anti-parallel $\beta$-sheets. In Fig.~\ref{fig6}, however, we observe 
a significant increase of the eigenvalue with $L$, which is not predicted.
The reason for the increase of the bend-mode eigenvalue 
with strand length $L$ is likely to be the significant 
bending of individual strands associated with this mode, which would
contribute a term to $\lambda_{\rm bend}$ that scales as $L^{4}$.

For the twist mode, as shown in Fig.~\ref{fig4}(a), we assume that
each strand rotates by an angle $\delta \theta$ with respect to its
neighboring strand. This corresponds to a uniform twist of the sheet
about the in-plane axis perpendicular to the strand orientation.  The
displacement of a C$_\alpha$ atom on the strand at $x$ along this
axis, and at a distance $l$ from this axis, is $\delta z \simeq l
(x/d) \delta \theta$, where $d$ is the distance between neighboring
strands.  It is straightforward to show that eigenvalue for the twist
mode goes as
\begin{equation}
\lambda_{\rm twist} \sim \langle \delta \theta^{2} \rangle L^{3} S^{3}
\end{equation}
At thermodynamic equilibrium, using $k_{B} T = c
\langle \delta \theta^{2} \rangle L S$, 
where $c$ is a twist stiffness per unit length, we find   
$\lambda_{\rm twist} \sim L^{2} S^{2}$. 
As shown in Fig.~\ref{fig6}, the twist-mode eigenvalue 
does scale as $L^2$ as predicted, at least for sheets of up 
to $S=5$ strands.  As seen in Fig.~\ref{fig5}, however, the scaling 
of the twist-mode eigenvalue with strand number is much weaker than
the predicted $S^2$, for all strand lengths.

\section{Discussion}

Our principal-component analysis of $\beta$-sheets indicates that
sheets in proteins are deformed primarily in two ways, by twisting and
bending. The amplitudes of these modes were found to have independent
Gaussian distributions, suggesting that twist and bend are the
``soft'' elastic modes of sheets.  The interpretation of the PCA twist
and bend modes as elastic modes of sheets is consistent with previous
PCA results on $\alpha$-helices. For helices, the dominant modes found
by PCA were shown to be indistinguishable from the soft elastic modes
of a model helix.\cite{Emberly}

In this light, the generally larger twist and bend deformations found
for anti-parallel sheets compared to parallel sheets suggests that
anti-parallel sheets have softer elastic spring constants. This could
arise from boundary effects imposed by the differences in
connectivity. Anti-parallel sheets tend to be connected by short loops
which allows the sheet to be easily bent. In contrast, parallel sheets
have  more complex connectivity (long loops) which could impinge on
their ability to deform. The stiffness may also arise due to slight
differences in the hydrogen bonding pattern between parallel and
sheared anti-parallel sheets\cite{Ho02}. It will be interesting to see if
physical modeling can capture the apparent differences in elasticity
between the two types of sheets.

While the PCA modes of all $\beta$-sheets studied had the
characteristics of elastic normal modes ({\it i.e.}, independent
Gaussian distributions of amplitudes) significant deviations from
simple scaling with sheet size were observed. These deviations can
probably be attributed to two main causes.  First, the actual soft
modes are not the pure twist and bend modes of rigid strands as
assumed in the scaling analysis. Significant deformations do occur
within individual strands - the bend mode also contains bending along
the strands which would contribute to the scaling of its eigenvalue
with strand length $L$. Second, $\beta$-sheets, as
quasi-two-dimensional objects, are subject to strong boundary
effects. For example, the constraint of global connectivity of the
sheets and the need to form backbone hydrogen bonds to the outer
strands are likely to introduce size-dependent effects beyond the
simple scaling analysis.  In contrast, the nearly ideal scaling of the
dominant modes of $\alpha$-helices probably reflects the much weaker
effect of boundaries on quasi-one-dimensional objects.\cite{Emberly}

Our PCA results provide a set of scaling behaviors
and structural properties that can be used to test and refine
energetic models of $\beta$-sheets. We have found that a simple spring
model, similar to one that was able to capture the normal modes of
helices\cite{Emberly}, did not describe well the scaling properties of
sheets. This suggests that the energetics governing $\beta$-sheet
flexibility is more complex than in helices.  More detailed force
fields are required and these will help to further clarify the
microscopic interactions governing $\beta$-sheet structure and
flexibility.

Another possible application of the PCA results is to
protein design.  A recent design scheme focussed on the packing of a
fixed set of structural elements to explore the space of potential
novel folds\cite{EmberlyPNAS}. In that work, only rigid helices were
considered for the structural building blocks. Because of the greater
flexibility of sheets compared to helices, extending the packing
scheme to sheets will require including energies of elastic
deformation. The soft modes of sheets can be easily incorporated into
the design of low-energy backbone structures using the elastic
energies $E(a_q) = c_q a_q^2/2$. Since we found only two dominant
modes for sheets, elasticity can be added to the packing scheme with
only two extra degrees of freedom per sheet.  The soft modes could
also be incorporated into models that analyze how a specific protein's
structure is changed when the sequence is mutated or redesigned.
Inclusion of soft modes may prove particularly useful in redesign of
binding sites, which is currently limited to the rigid-backbone
approximation\cite{looger}.

In summary, this work has used database protein
structures to reveal the flexible motions of $\beta$-sheets. The
effective spring constants and eigenvectors for the sheets studied, as
well as the mean structure coordinates, are available as Supplementary
Material.

\section{Methods}
We compiled a set of $\beta$-sheet structures from $2860$
representative protein folds in the FSSP database\cite{FSSP}.  The
representative structures in the FSSP are the structurally distinct
folds that result from doing an all-against-all structure clustering
of protein folds that have less than 25\% sequence identity. Thus the
set is designed to minimize fold redundancy. Making use of the
structural annotations in the PDB files for the representative set, we
extracted the $\beta$-sheets from each fold.  Only sheets within a
certain size range were considered: sheets had to have at least 3
strands and no more than 25, with between 3 and 15 residues per
strand.  Using these criteria we were able to extract $3516$
representative $\beta$-sheets.

Many of the sheets in this representative set were found to have defects.
A defect occurs when there is a gap in the hydrogen-bonding pattern,
or the strands are of different lengths and one overhangs the other. 
These two types of defects are schematically illustrated in 
Fig.~\ref{fig8}. We wished to eliminate sheets with defects before
performing a principal-component analysis of flexibility. To 
systematically identify defects, we developed a procedure for
finding the optimal pairing of C$_\alpha$ atoms on adjacent strands. 
A defect is indicated when a C$_\alpha$ atom is left
unpaired. To find the optimal pairing, we computed a distance 
matrix $d_{i,j}$ for each pair of neighboring strands, 
\begin{equation}
d_{i,j} = |\vec{r}_i - \vec{r}_j|
\label{eq:sij}
\end{equation}
where $\vec{r}_i$ is the position of the $i$th C$_\alpha$ atom on
the first strand and $\vec{r}_j$ is the position of the $j$th C$_\alpha$ 
atom on the second strand. The two strands could be of different
lengths. We defined the optimal pairing $i \Leftrightarrow j$ between 
strands to be the one which minimized 
\begin{equation}
D =  6 \AA N_{\rm gaps} + \sum_{{\rm pairs} i,j} d_{i,j}
\label{eq:sumsij}
\end{equation}
where the last term represents a penalty of +6\AA\ for each gap.  A
gap occurs when a $C_\alpha$ atom on one strand is not paired.  In
practice, we found that a gap penalty of 6\AA\ identified gaps in good
agreement with a visual inspection of the sheets.  To efficiently find
the optimal pairing, {\it i.e.} the one which minimizes $D$, we
employed the Needleman-Wunsch method\cite{Needleman} for global
sequence alignment. We applied the alignment procedure for each
neighboring pair of strands in the sheet (strands in the sheet tend to
be annotated in the order they occur in the sheet, hence the need to
only align neighbors). This gave us the optimal alignment of all the
strands in the sheet, and allowed us to identify sheet defects.

After performing the pairwise alignment of all the strands making
up a sheet we then extracted sheets of a fixed size, {\it e.g.} 
$S$ strands each of length $L$. This is illustrated in
Fig.~\ref{fig8}(a). We scanned the aligned sheet using a window of the
specified dimension. If the window spanned a region that did not contain
a gap, the positions of the C$_\alpha$ atoms were recorded to a file.
For each extracted sheet, we also recorded the sheet geometry, 
{\it i.e.} the relative amino-terminal to carboxyl-terminal directions 
of all the strands in the sheet. The direction of the first strand was 
defined to be up to simplify the recording of sheet geometries. 
The choice of which of the two outermost strands to call the first 
strand was arbitrary, and this was compensated for later.
Another quantity recorded for each sheet was
its ``pleatedness''. A $\beta$-sheet is corrugated - it has alternating 
ridges and valleys. If the first row of atoms from the
strands making up a sheet comprise a ``ridge'' we consider the sheet to
have positive pleatedness. If the first row of atoms comprise a 
``valley'' we consider the sheet to
have negative pleatedness. The two types of pleatedness are shown in
Fig.~\ref{fig8}(b).  We consider sheets of the same size, geometry, 
and pleatedness to constitute a ``class''. Only sheets of the same class 
are aligned for subsequent principal-component analysis (see Results).

We note that each extracted ungapped sheet in fact generates two
sheets because either of the two outermost strands could be called
the ``first''. Having arbitrarily chosen a first strand and 
called its direction up, there are two possible symmetry operations 
to obtain the second sheet. If the last strand is also up, we flip 
the sheet (this causes the pleatedness to change sign).
If on the other hand the last strand is oriented down, we perform a clockwise 
rotation of the sheet by 180$^o$. These symmetry operations effectively 
double our database of sheet structures.

\begin{figure}[!t]
\caption{Distribution of geometries for $\beta$-sheets containing at 
least 6 strands from the set of $3516$ representative structures. 
The strand orientations are labeled with ``u'' for up and ``d'' for 
down, according to amino-terminal to carboxyl-terminal direction.
We have fixed the orientation of the first strand to always be
up.\label{fig1}
}
\end{figure}

\begin{figure}[!t]
\caption{Demonstration set of $10$ anti-parallel sheets,
of 4 strands each with 5 residues per strand and the same
pleatedness (corrugation), all aligned to the average sheet 
for this class.\label{fig2}
}
\end{figure}

\begin{figure}[!t]
\caption{The ten largest eigenvalues from the principal-component
analysis of $1454$ anti-parallel 4 $\times$ 5 $\beta$-sheets, of the
class shown in Fig.~\protect\ref{fig2}
\label{fig3}
}
\end{figure}

\begin{figure}[!t]
\caption{(a) Exaggerated twist mode of a 6 stranded $\beta$-sheet
with 5 residues per strand. (b) Exaggerated bend mode of the 
same sheet.  Average structures are shown in gray, deformed structures 
are shown in red. 
\label{fig4}}
\end{figure}

\begin{figure}[!t]
\caption{Scaling of the eigenvalues for the bend and twist modes as a
function of the number of strands for sheets of fixed strand length.
The scalings are shown for both parallel and anti-parallel sheets. 
\label{fig5}}
\end{figure}

\begin{figure}[!t]
\caption{Scaling of the eigenvalues for the bend and twist modes as a
function of strand length for sheets with fixed number of strands.
The scalings are shown for both parallel and anti-parallel sheets. 
\label{fig6}}
\end{figure}

\begin{figure}[!t]
\caption{(a,b) Distribution of the projections of sheet displacement
onto the twist and bend modes for $1454$ anti-parallel $\beta$-sheets
each with 4 strands of 5 residues (cf. Fig.~\protect\ref{fig2}).
Gaussian fits to the distributions are shown by
the red curves.  (c) Projections onto the
subspace spanned by the twist and bend modes for the same 1454 structures.
\label{fig7} }
\end{figure}

\begin{figure}[!t] 
\caption{
(a) Schematic of the alignment procedure used to extract 
$\beta$-sheets of a fixed dimension. On the left is sheet that might exist 
within the database.  Backbone C$_\alpha$ atoms are aligned
(dashed lines) using a global alignment procedure that allows
for gaps. For example, a gap occurs in the bottom strand at the second 
C$_\alpha$ atom from the left. All ungapped
sheets of given dimension ($S$ strands each of length $L$) are then
extracted. (b)
The two types of pleatedness. If the first (leftmost) row of atoms forms
a ridge, we consider the sheet to
have positive pleatedness.  If the first row of atoms forms a valley 
we consider the sheet to have negative pleatedness.
\label{fig8}}
\end{figure}









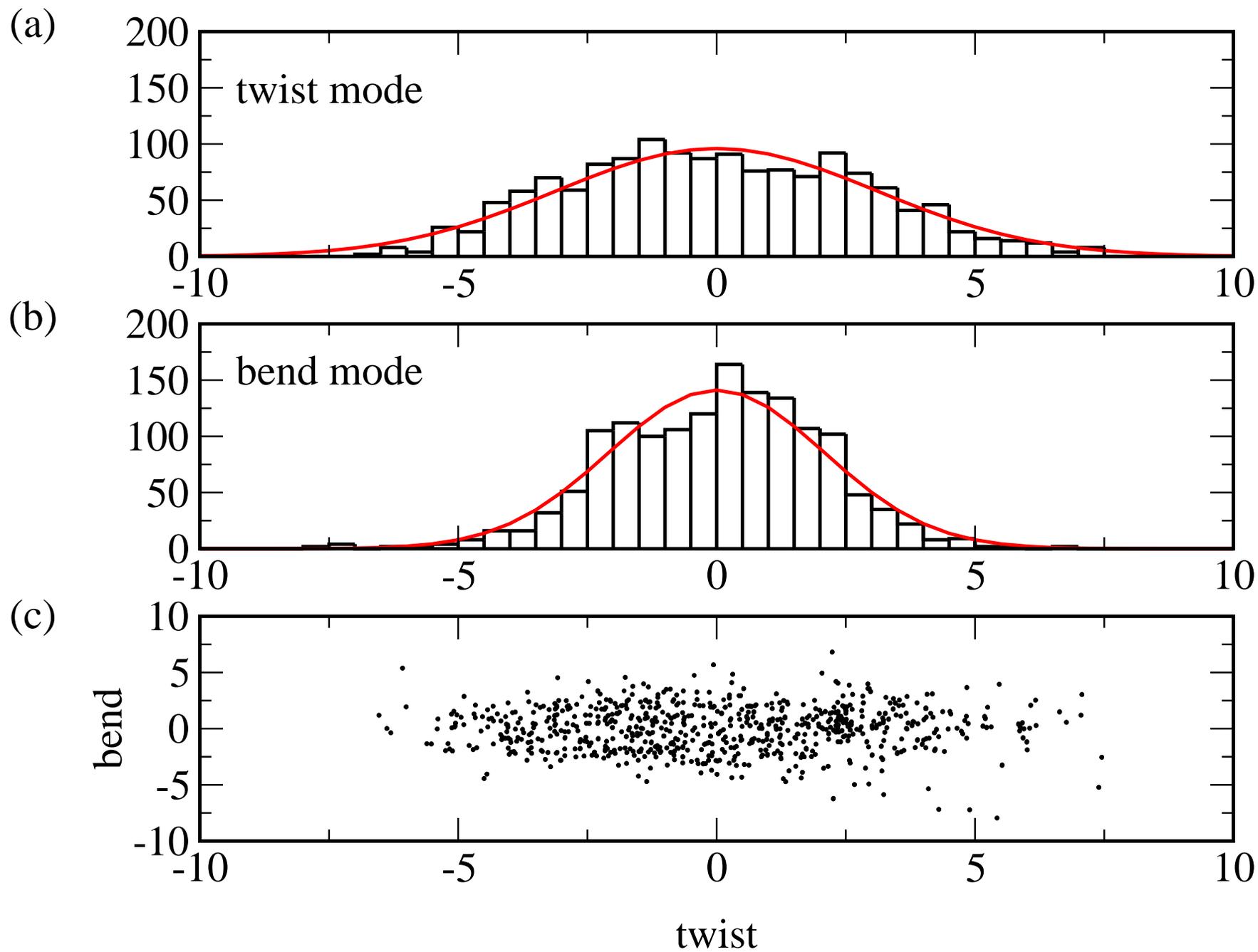
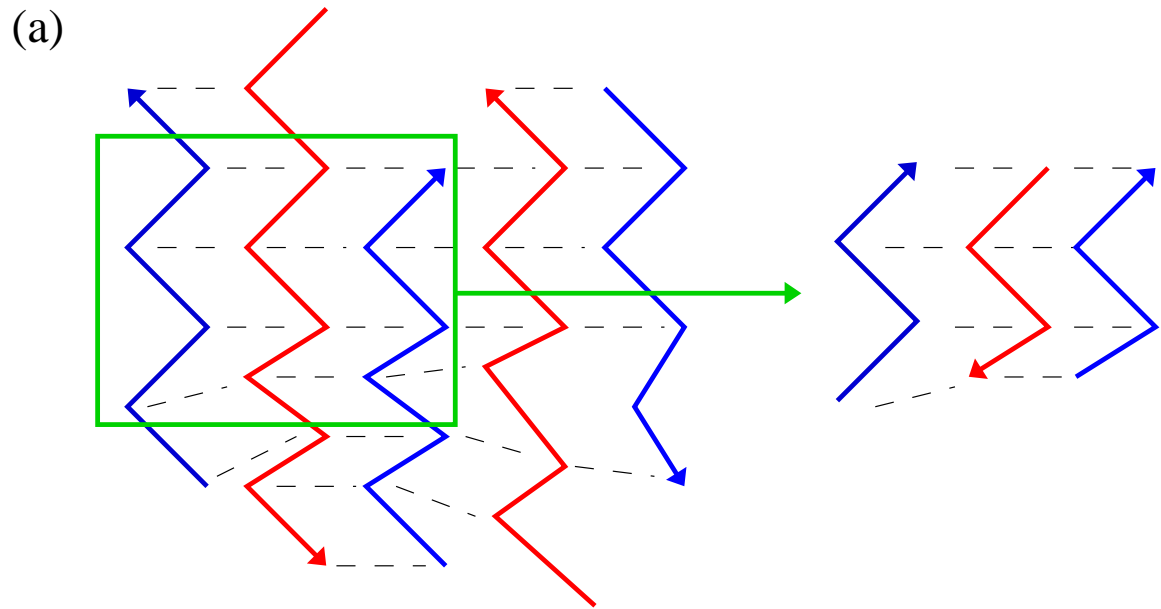
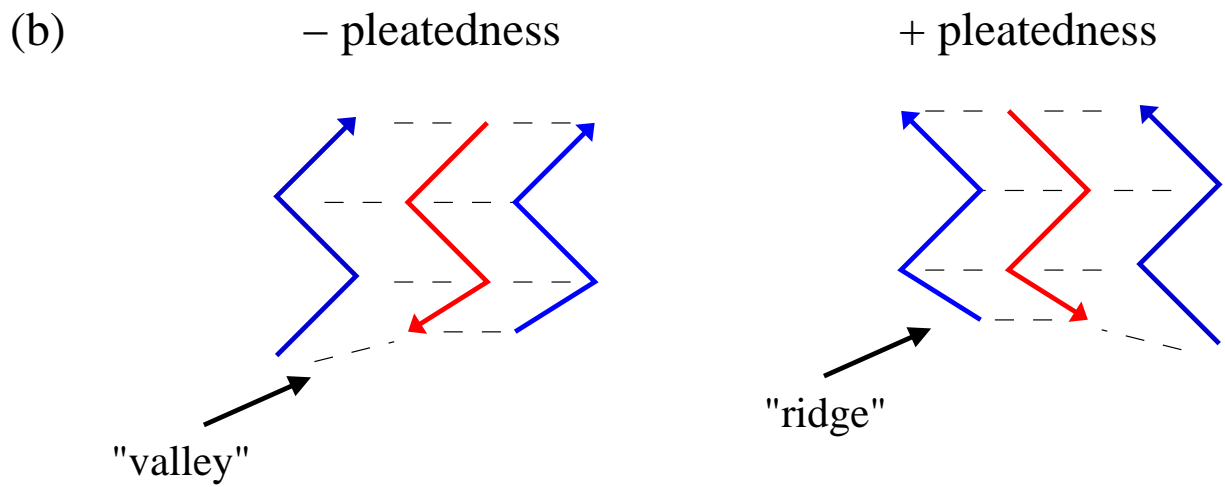
\end{document}